\documentclass[aps,11pt]{revtex4}
\usepackage{dcolumn}
\usepackage{graphicx}
\usepackage{amsmath}
\usepackage{amsfonts}
\usepackage{amssymb}
\usepackage{psfrag}
\usepackage{wrapfig}
\usepackage{subfigure}
\usepackage{makeidx}
\usepackage{bm}
\usepackage{epsf}
\usepackage{hyperref}
\usepackage{color}
\usepackage{cases}
\makeatletter

\newcommand{\Rmnum}[1]{\expandafter\@slowromancap\romannumeral #1@}
\newcommand{\eq}[1]{Eq.(\ref{#1})}
\makeatother

\begin{document}

\title{Analytical approximate solutions of AdS black holes in Einstein-Weyl-scalar gravity}

\author{Ming Zhang$^{1,2}$\footnote{e-mail: mingzhang0807@126.com; zhangming@xaau.edu.cn}, Sheng-Yuan Li$^{3}$\footnote{e-mail: lishengyuan314159@hotmail.com}, De-Cheng Zou$^{4,3}$\footnote{Corresponding author: dczou@jxnu.edu.cn}, Chao-Ming Zhang$^{3}$\footnote{e-mail: zcm843395448@163.com} }

\affiliation{$^{1}$Faculty of Science, Xi'an Aeronautical University, Xi'an 710077 China\\
$^{2}$School of Physics, Northwest University, Xi'an, 710069, China\\
$^{3}$Center for Gravitation and Cosmology, College of Physical Science and Technology, Yangzhou University, Yangzhou 225009, China\\
$^{4}$College of Physics and Communication Electronics, Jiangxi Normal University, Nanchang 330022, China}

\date{\today}

\begin{abstract}
\indent

We consider Einstein-Weyl gravity with a minimally coupled scalar field in four dimensional spacetime. Using the minimal geometric deformation (MGD) approach, we split the highly nonlinear coupled field equations into two subsystems that describe the background geometry and scalar field source, respectively. By considering the Schwarzschild-AdS metric as background geometry, we derive analytical approximate solutions of the scalar field and deformation metric functions using the homotopy analysis method (HAM), providing their analytical approximations to fourth order. Moreover, we discuss the accuracy of the analytical approximations, showing they are sufficiently accurate throughout the exterior spacetime.

\end{abstract}


\maketitle

We consider Einstein-Weyl gravity with a minimally coupled scalar field in four dimensional spacetime. Using the minimal  geometric deformation (MGD) approach, we split the highly nonlinear coupled field equations into two subsystems that describe the background geometry and scalar field source, respectively. By considering the Schwarzschild-AdS metric as background geometry, we derive analytical approximate solutions of the scalar field and deformation metric functions using the homotopy analysis method (HAM), providing their analytical approximations to fourth order. Moreover, we discuss the accuracy of the analytical approximations, showing they are sufficiently accurate throughout the exterior spacetime.

\section{Introduction}
\label{intro}

Despite the comprehensive and highly precise experimental testing of general relativity (GR) thus far, it is recognized
as a non-renormalizable quantum field theory from a theoretical perspective. One potential solution to this challenge involves introducing  higher-order corrections that gain significance at higher energy\cite{Stelle:1976gc}.
In four dimensional spacetime, the most general higher derivative theory of gravity is Einstein gravity
with an additional second order in the curvature term, which has the following form\cite{Holdom:2016nek,Lu:2015cqa}:
\begin{equation}
\mathcal{I} =\int d^4 x\sqrt{-g}\left( \gamma R-\beta C_{\mu\rho\nu\sigma} C^{\mu\rho\nu\sigma}+\delta R^2 \right),\label{Action1}
\end{equation}
where the parameters $\gamma$, $\beta$, and $\delta$ are constants, $R$ is the Ricci scalar, and $C_{\mu\rho\nu\sigma}$ is the Weyl tensor.
In gravitational theory, black holes hold a pivotal position as the most fundamental objects and are also believed to provide an important window to the quantum nature of gravity. In pure Einstein-Weyl gravity,
non-Schwarzschild black hole (NSBH) solutions have been obtained in four dimensional \cite{Lu:2015cqa,Lu:2015psa,Kokkotas:2017zwt}
and higher dimensional~\cite{Lu:2017kzi} spacetime, even generalizations of AdS~\cite{Podolsky:2018pfe,Svarc:2018coe}
and charged solutions~\cite{Lin:2016jjl,Wu:2019uvq}. In addition,Refs. \cite{Konoplya:2019ppy,Cai:2015fia,Zinhailo:2018ska} discuss the
Hawking radiation in the vicinity of NSBHs and the quasinormal modes under test scalar field perturbation.

In recent years, a novel and straightforward approach known as gravitational decoupling~\cite{Ovalle:2017fgl} has emerged as a promising method for separating gravitational
sources within the framework of GR. The minimal geometric deformation (MGD) technique~\cite{Ovalle:2016pwp}
is an important method in the theory of gravitational decoupling. Originally introduced within the context of the Randall-Sundrum braneworld~\cite{Randall:1999ee,Randall:1999vf}, the MGD technique has been successfully employed to generate brane-world~\cite{Ovalle:2020fuo,Ovalle:2007bn,Ovalle:2009xk} configurations
from initial solutions based on general relativistic perfect fluids. Building upon its initial application, this technique offers the potential to obtain static and spherically symmetric solutions with more realistic sources than the ideal perfect fluid model and has been further extended to construct physically meaningful interior stellar solutions~\cite{Casadio:2015gea,Ovalle:2015nfa,Ovalle:2013xla}. In Ref.\cite{Ovalle:2018umz}, the MGD technique was also used to calculate the deformation of black hole solutions under different conditions
in a asymptotically flat Schwarzschild black hole system. Notably, recent studies have demonstrated the applicability of gravitational decoupling
in alternative theories of gravity, such as $f(R, T)$ gravity, Gauss-Bonnet gravity, $f(G)$ gravity, and Rastall gravity \cite{Vacaru:2011ydi,Maurya:2020ebd,Sharif:2020llo,Sharif:2020rlt,Maurya:2019xcx}. In addition to static spacetime,
the MGD technique has also found excellent applications in rotating spacetime~\cite{Contreras:2021yxe}, and in  de-Sitter spacetime,
this technique has been used for the deformation of black hole solutions \cite{Gabbanelli:2019txr}. These advancements highlight
the versatility and wide-ranging implications of the MGD technique in various gravitational theories.

Inspired by these, we focus on black hole solutions in Einstein-Weyl gravity with a minimally coupled scalar field. Using the MGD technique, we start with a
seed metric for a vacuum solution in Einstein-Weyl gravity and construct black hole solutions with more complex forms of the energy momentum tensor
for scalar field. However, obtaining new black hole solutions is difficult in general, owing to the complicated nonlinear field equations for metric functions and scalar fields. The homotopy analysis method (HAM) proposed by Liao \cite{Liao:1992mua,Liao:2003mua,Liao2004} has gained significant traction in the past decade as a powerful tool for studying a wide range of mathematical and physical problems\cite{van2008,abb2008,saj2008,chen2009,tur2017,van2017}. However, in the context of gravitational theories, the application and research of the HAM remain noticeably scarce \cite{Sultana:2019lhf,Sultana:2021cvq,Zou:2023inv}. Here, we adopt the HAM to derive analytical approximate solutions
for hairy black holes in Einstein-Weyl-scalar gravity.

This paper is organized as follows. In Sec. \ref{se2}, we discuss AdS black holes in Einstein-Weyl-scalar gravity and derive the corresponding field equations
using the MGD technique. In Sec. \ref{aas}, we briefly introduced the HAM and derived the analytical approximate
solutions for the system. Finally, the paper concludes with a discussion on the results obtained in Sec. \ref{conc}.

\section{ Einstein-Weyl AdS black holes with \textbf{a} minimally coupled scalar field}\label{se2}

We take a minimally coupled scalar field $\phi$ as the source, such that the action for AdS black holes in Einstein-Weyl-scalar gravity is given by
\begin{eqnarray}
  I=\int{d^4x \sqrt{-g}(R-2\Lambda-\beta C_{\mu\rho\nu\sigma}C^{\mu\rho\nu\sigma}
  -\frac{1}{2}\partial_{\mu}\phi\partial^{\mu}\phi-V(\phi))}.
\end{eqnarray}
where $\Lambda$ is the cosmological constant, and $V$ the potential function.
The corresponding field equations with respect to the field variables $g_{\mu\nu}$ and $\phi$ are obtained as
\begin{eqnarray}
&& E_{\mu\nu}=R_{\mu\nu}-\frac{1}{2}g_{\mu\nu}R-\Lambda g_{\mu\nu}-4\beta B_{\mu\nu}=T_{\mu\nu}^\phi \label{eom},\\
&&\Box\phi-\frac{dV}{d\phi}=0,\label{KG}
\end{eqnarray}
where $\Box\phi=g^{\mu\nu}\phi_{;\mu\nu}$, $T_{\mu\nu}^\phi$ is the energy momentum tensor of scalar field $\phi$,
\begin{eqnarray}
  T_{\mu\nu}^\phi=
  \nabla_{\mu}\phi\nabla_{\nu}\phi-\frac{1}{2}g_{\mu\nu}\nabla^{\alpha}\phi
  \nabla_{\alpha}\phi-V(\phi)g_{\mu\nu}.
\end{eqnarray}

and the $B_{\mu\nu}$ denotes the trace-free Bach tensor with
\begin{eqnarray}
  B_{\mu\nu}=(\nabla^{\rho}\nabla^{\sigma}+\frac{1}{2}R^{\rho\sigma})
  C_{\mu\rho\nu\sigma}.
\end{eqnarray}

Now, we assume the static and spherically symmetric metric
\begin{eqnarray}
  ds^2=-h(r)dt^2+\frac{dr^2}{\xi(r)}+r^2(d\theta^2+\sin^2\theta d\varphi^2).
\end{eqnarray}
Substituting the above metric ansatz into the field equations \eqref{eom}, and \eqref{KG}, we can obtain highly nonlinear and coupled field equations. In general, derive black hole solutions for the functions $h(r)$ and $\xi(r)$ is difficult owing to the complicated forms of Eqs. \eqref{eom} and \eqref{KG}.

To address this challenge, Ovalle et al. \cite{Ovalle:2017fgl, Ovalle:2016pwp} presented the so-called MGD method, which states that the metric function $\xi(r)$ can be separated into
\begin{eqnarray}
  \xi(r)\rightarrow f(r)+\alpha \mu(r) \label{fmu},
\end{eqnarray}

which leads to a new form of the gravitational field equation \eqref{eom},
\begin{eqnarray}
&&\bar{E}_{\mu\nu}+\alpha\tilde{E}_{\mu\nu}=T_{\mu\nu}^\phi.
\end{eqnarray}

Here, $\alpha$ is a decoupling constant, and the function $f(r)$
 corresponds to the vacuum solution of the field equation of Einstein-Weyl-AdS gravity,
\begin{eqnarray}
 \bar{E}_{\mu\nu}=0,~~\text{to find}~~\{h(r), f(r) \}\label{eqe1}.
\end{eqnarray}


Then, we can obtain
\begin{eqnarray}
rh[rf'h'+2f(rh''+2h')]+4h^2(rf'+f-1+2\Lambda r^2)-r^2fh'^2=0, \label{neom1}\\
f''+\frac{r^2fh'^2+2rfhh'+4(f-1+2\Lambda r^2)h^2}{2rfh(rh'-2h)}f'-\frac{3hf'^2}{4fh-2rfh'}+8\Lambda
\frac{h-fh-rfh'-\Lambda r^2h}{3f(2h-rh')} \nonumber\\
-\frac{r^3fh'+(r^2f-r^2+\Lambda r^4)h}{\beta r^2f(rh'-2h)}-\frac{r^3fh'^3-3r^2fhh'^2
-8(f-1)h^3}{2r^2h^2(rh'-2h)}=0. \label{neom2}
\end{eqnarray}

Fortunately, Lin et al. \cite{Lin:2016kip} presented two groups of black hole solutions for the functions $f(r)$ and $h(r)$, the Schwarzschild AdS black hole solutions ($h(r)=f(r)$) and non-Schwarzschild AdS black hole solutions ($h(r)\neq f(r)$).

In addition, the function $\mu(r)$ denotes the contributions from the energy-momentum tensor  $T_{\mu\nu}^\phi$ of the scalar field $\phi$ and satisfies the quasi-gravitational field equation
\begin{eqnarray}    \alpha\tilde{E}_{\mu\nu}=T_{\mu\nu}^\phi. \label{eqe2}
\end{eqnarray}

With the scalar field equation \eqref{KG} and quasi-gravitational field equation \eqref{eqe2}, we can obtain
\allowdisplaybreaks[1]
\begin{eqnarray}
  \phi''(r)=\frac{-h r f' \phi '^2-f r h' \phi '^2-4 f h \phi '^2-\alpha  \mu  r h' \phi '^2-4 \alpha  h \mu  \phi '^2-\alpha  h r \mu ' \phi '^2+2 h r V'}{2rh\phi' (f+\alpha\mu)}, \label{eqphi}
\end{eqnarray}
\begin{eqnarray}
  \mu''(r)&=&\Big(-32 h^3 \alpha ^2 \beta  \Lambda ^2 \mu ^2 h' r^5+12 h^3 \alpha ^2 \Lambda  \mu ^2 h' r^5-32 f h^3 \alpha  \beta  \Lambda ^2 \mu  h' r^5+12 f h^3 \alpha  \Lambda  \mu  h' r^5 \nonumber\\
  &+&6 f^2 \alpha ^2 \beta  \mu ^2 h'^4 r^4+6 f^3 \alpha  \beta  \mu  h'^4 r^4-12 f h \alpha ^2 \beta  \mu ^2 f' h'^3 r^4-12 f^2 h \alpha  \beta  \mu  f' h'^3 r^4 \nonumber\\
  &+&64 h^4 \alpha ^2 \beta  \Lambda ^2 \mu ^2 r^4+48 h^4 \alpha ^2 \Lambda  \mu ^2 r^4+12 f h^2 \alpha ^2 \mu ^2 h'^2 r^4-32 f h^2 \alpha ^2 \beta  \Lambda  \mu ^2 h'^2 r^4 \nonumber\\
  &-&15 h^2 \alpha ^2 \beta  \mu ^2 f'^2 h'^2 r^4-18 f h^2 \alpha  \beta  \mu  f'^2 h'^2 r^4+12 f^2 h^2 \alpha  \mu  h'^2 r^4-32 f^2 h^2 \alpha  \beta  \Lambda  \mu  h'^2 r^4  \nonumber\\
  &+&3 f^2 h^2 \alpha ^2 \beta  h'^2 \mu '^2 r^4+108 f^3 h^4 \phi '^2 r^4+108 f^2 h^4 \alpha  \mu  \phi '^2 r^4+64 f h^4 \alpha  \beta  \Lambda ^2 \mu  r^4 \nonumber\\
  &-&168 f h^4 \alpha  \Lambda  \mu  r^4+216 f^2 h^4 V(r) r^4-144 h^3 \alpha ^2 \beta  \Lambda  \mu ^2 f' h' r^4-36 f h^3 \alpha  \mu  f' h' r^4  \nonumber\\
  &-&144 f h^3 \alpha  \beta  \Lambda  \mu  f' h' r^4+6 f^3 h \alpha  \beta  h'^3 \mu ' r^4+6 f^2 h \alpha ^2 \beta  \mu  h'^3 \mu ' r^4+18 f^2 h^2 \alpha  \beta  f' h'^2 \mu ' r^4  \nonumber\\
  &+&12 f h^2 \alpha ^2 \beta  \mu  f' h'^2 \mu ' r^4+36 f^2 h^3 \alpha  h' \mu ' r^4+96 f^2 h^3 \alpha  \beta  \Lambda  h' \mu ' r^4+96 f h^3 \alpha ^2 \beta  \Lambda  \mu  h' \mu ' r^4 \nonumber\\
  &-&30 f^2 h \alpha ^2 \beta  \mu ^2 h'^3 r^3-30 f^3 h \alpha  \beta  \mu  h'^3 r^3-24 f h^2 \alpha ^2 \beta  \mu ^2 f' h'^2 r^3-24 f^2 h^2 \alpha  \beta  \mu  f' h'^2 r^3  \nonumber\\
  &-&12 f^2 h^3 \alpha ^2 \beta  h' \mu '^2 r^3+288 h^4 \alpha ^2 \beta  \Lambda  \mu ^2 f' r^3-144 f h^4 \alpha  \mu  f' r^3+288 f h^4 \alpha  \beta  \Lambda  \mu  f' r^3  \nonumber\\
  &+&60 f h^3 \alpha ^2 \mu ^2 h' r^3-12 h^3 \alpha ^2 \mu ^2 h' r^3+32 f h^3 \alpha ^2 \beta  \Lambda  \mu ^2 h' r^3+32 h^3 \alpha ^2 \beta  \Lambda  \mu ^2 h' r^3  \nonumber\\
  &-&30 h^3 \alpha ^2 \beta  \mu ^2 f'^2 h' r^3-18 f h^3 \alpha  \beta  \mu  f'^2 h' r^3+60 f^2 h^3 \alpha  \mu  h' r^3-12 f h^3 \alpha  \mu  h' r^3  \nonumber\\
  &+&32 f^2 h^3 \alpha  \beta  \Lambda  \mu  h' r^3+32 f h^3 \alpha  \beta  \Lambda  \mu  h' r^3+24 f^3 h^2 \alpha  \beta  h'^2 \mu ' r^3+24 f^2 h^2 \alpha ^2 \beta  \mu  h'^2 \mu ' r^3  \nonumber\\
  &+&144 f^2 h^4 \alpha  \mu ' r^3-192 f^2 h^4 \alpha  \beta  \Lambda  \mu ' r^3-192 f h^4 \alpha ^2 \beta  \Lambda  \mu  \mu ' r^3+24 f h^3 \alpha ^2 \beta  \mu  f' h' \mu ' r^3  \nonumber\\
  &+&48 f h^4 \alpha ^2 \mu ^2 r^2-48 h^4 \alpha ^2 \mu ^2 r^2+64 f h^4 \alpha ^2 \beta  \Lambda  \mu ^2 r^2-64 h^4 \alpha ^2 \beta  \Lambda  \mu ^2 r^2+120 h^4 \alpha ^2 \beta  \mu ^2 f'^2 r^2  \nonumber\\
  &+&108 f h^4 \alpha  \beta  \mu  f'^2 r^2+36 f^2 h^2 \alpha ^2 \beta  \mu ^2 h'^2 r^2+36 f^3 h^2 \alpha  \beta  \mu  h'^2 r^2+12 f^2 h^4 \alpha ^2 \beta  \mu '^2 r^2  \nonumber\\
  &+&48 f^2 h^4 \alpha  \mu  r^2+24 f h^4 \alpha  \mu  r^2+64 f^2 h^4 \alpha  \beta  \Lambda  \mu  r^2-64 f h^4 \alpha  \beta  \Lambda  \mu  r^2+24 f h^3 \alpha ^2 \beta  \mu ^2 f' h' r^2  \nonumber\\
  &+&72 h^3 \alpha ^2 \beta  \mu ^2 f' h' r^2+24 f^2 h^3 \alpha  \beta  \mu  f' h' r^2+72 f h^3 \alpha  \beta  \mu  f' h' r^2-72 f^2 h^4 \alpha  \beta  f' \mu ' r^2  \nonumber\\
  &-&96 f h^4 \alpha ^2 \beta  \mu  f' \mu ' r^2-24 f^3 h^3 \alpha  \beta  h' \mu ' r^2-48 f^2 h^3 \alpha  \beta  h' \mu ' r^2-24 f^2 h^3 \alpha ^2 \beta  \mu  h' \mu ' r^2  \nonumber\\
  &-&48 f h^3 \alpha ^2 \beta  \mu  h' \mu ' r^2+144 f h^4 \alpha ^2 \beta  \mu ^2 f' r-144 h^4 \alpha ^2 \beta  \mu ^2 f' r+144 f^2 h^4 \alpha  \beta  \mu  f' r  \nonumber\\
  &-&144 f h^4 \alpha  \beta  \mu  f' r-48 f^2 h^3 \alpha ^2 \beta  \mu ^2 h' r+48 f h^3 \alpha ^2 \beta  \mu ^2 h' r-48 f^3 h^3 \alpha  \beta  \mu  h' r  \nonumber\\
  &+&48 f^2 h^3 \alpha  \beta  \mu  h' r-96 f^3 h^4 \alpha  \beta  \mu ' r+96 f^2 h^4 \alpha  \beta  \mu ' r-96 f^2 h^4 \alpha ^2 \beta  \mu  \mu ' r+96 f h^4 \alpha ^2 \beta  \mu  \mu ' r  \nonumber\\
  &+&96 f^2 h^4 \alpha ^2 \beta  \mu ^2-96 f h^4 \alpha ^2 \beta  \mu ^2+48 h^4 \alpha ^2 \beta  \mu ^2+96 f^3 h^4 \alpha  \beta  \mu -96 f^2 h^4 \alpha  \beta  \mu   \nonumber\\
  &+&96 f h^4 \alpha  \beta  \mu \Big)\Big/\Big(12r^2\alpha\beta f^2h^2(f+\alpha\mu)(-2h+rh')^2 \Big) \label{eqmu}
\end{eqnarray}
and
\begin{eqnarray}
  V(r)=\frac{\alpha  r^2 h' \left(\mu  f'-f \mu '\right)+h \left(2 \alpha  \mu  \left(2 r f'-f r^2 \phi'^2+4 \Lambda  r^2-2\right)-2 f r \left(2 \alpha  \mu '+f r \phi'^2\right)\right)}{8 f h r^2} .\label{eqv}
\end{eqnarray}

Clearly, it is a challenging task to find exact solutions for the deformation function $\mu(r)$, scalar field $\phi(r)$, and potential $V(\phi)$ from the nonlinear differential equations \eqref{eqphi}, \eqref{eqmu}, and \eqref{eqv}, because the field equations  \eqref{eqphi}, \eqref{eqmu}, and \eqref{eqv} also refer to the metric functions $h(r)$ and $f(r)$. For simplicity, we can adopt the Schwarzschild AdS metric $[h(r)=f(r)=1-\frac{2M}{r}-\frac{\Lambda}{3}r^2]$ as the seed solution to derive  solutions for scalarized AdS black holes. Moreover, we employ the HAM to analytically derive these solutions in next Section.

\section{\textbf{Analytical approximate} solutions}
\label{aas}

 In this section, we apply the HAM to derive analytical approximate solutions for the deformation function $\mu(r)$, scalar field $\phi(r)$, and potential $V(\phi)$.

Consider an $n$-nonlinear differential equation system, where $y_i(t)$ is the solution of the nonlinear operator $N_i$ as a function of $t$,
\begin{eqnarray}
  N_i[y_i(t)]=0, \qquad\quad i=1,2,...,n, \label{Nequation}
\end{eqnarray}
with an unknown function $y_i(t)$ and a variable $t$. Then, the zero-order deformation equation can be written as
\begin{eqnarray}
(1-q)L[\phi_i(t;q)-y_{i0}(t)]=q h_i H_i(t)N_i[\phi_i(t;q)].\label{hm}
\end{eqnarray}
The HAM involves constructing a topological homotopy that incorporates a linear auxiliary operator $L$ and nonlinear operator $N_i$. By introducing an embedding parameter $q\in[0,1]$, the solution of the entire equation undergoes a continuous transformation from the solution $y_{i0}(t)$ of the chosen linear auxiliary operator $L$ to the solution $y_i(t)$ of the nonlinear equation $N_i$ as $q$ varies from $0$ to $1$.
To ensure the convergence of the solution, an auxiliary function $H_i(t)$ and convergence control parameter $h_i$ are introduced into the homotopy equation \eqref{hm}. By carefully selecting an appropriate auxiliary function $H_i(t)$ and convergence control parameter $h_i$, the solution can converge more rapidly, enhancing the efficiency of the method.

To decompose the nonlinear problem into a series of linear subproblems, we perform Taylor expansions of $\phi_i(t;q)$ with respect to $q$ around $q = 0$
\begin{eqnarray}
\phi_i(t;q)=y_{i0}(t)+\sum_{m=1}^{\infty}y_{im}(t)q^m.\label{phiexpand}
\end{eqnarray}

Here, the coefficient $y_{im}(t)$ of the $m$-th order of $q$ is
\begin{eqnarray}
y_{im}(t)=\frac{1}{m!}\frac{\partial^m \phi_i(t;q)}{\partial q^m}.\label{coeffyim}
\end{eqnarray}

When $q = 1$, it is the expansion of the solution of the nonlinear equation (\ref{Nequation})\textbf{,}
\begin{eqnarray}
  y_i(t)=\phi_i(t;1)=y_{i0}(t)+\sum_{m=1}^{\infty}y_{im}(t).
\end{eqnarray}

The expansion of the solutions of the nonlinear equations can be found as long as $y_{im}(t)$ is solved. To achieve this, the operation for the zero order deformation equation (\eq{hm}) is as follows: First, substitute the expansion (\eq{phiexpand}) into \eq{hm}. Second, take the $m$-th derivative of $q$ on (\eq{hm}) both sides. Third, after calculating the derivatives, set $q=0$. Then, the so-called higher order deformation equation ($m$th-order deformation equation) is obtained, which is given by
\begin{eqnarray}
  L[y_{im}(t)-\chi_my_{im-1}(t)]=h_i H_i(t) R_{im}(y_{im-1}).
  \label{m-orderhm}
\end{eqnarray}
Where the term $R_{im}$ on the right-hand side with respect to the nonlinear operator is
\begin{eqnarray}
\label{Rim}
  R_{im}(y_{im-1})=\frac{1}{(m-1)!}\frac{\partial^{m-1} N_i[\sum_{m=0}^{\infty}y_{im}(t)q^m]}{\partial q^{m-1}}\mid_{q=0},
\end{eqnarray}
and the constant
\begin{eqnarray}
  \chi_m=\left\{
  \begin{aligned}
    0 & : m\leq 1 \\
    1 & : m>1
  \end{aligned}
  \right.
\end{eqnarray}
In \eq{Rim}, the highest term on the right-hand side can only reach up to the $y_{im-1}$ term. Therefore, according to \eq{m-orderhm}, $y_{im}(t)$ is related to $y_{im-1}$, allowing us to determine $y_{im}(t)$ of any desired order using this relationship.
In the calculation, we take a finite order to ensure that the error is sufficiently small, a finite M-order approximation,
\begin{eqnarray}
\label{m-y-expand}
  y_i^M(t)=y_{i0}(t)+\sum_{m=1}^{M}y_{im}(t),
\end{eqnarray}
where $y_i^M(t)$ is the M-th order approximate solution of the original equation (\ref{Nequation}).

The HAM offers significant flexibility in choosing the linear auxiliary operator $L$, initial guess $y_{i0}(t)$, auxiliary function $H_i(t)$, and convergence control parameter $h_i$, which makes it adaptable to a wide range of nonlinear problems. However, owing to this freedom of choice, it is essential to have a theoretical foundation to determine these quantities more appropriately. Further guidance and theoretical considerations for selecting these quantities can be found in Refs.\cite{Robert2009,Yinshan2010}.

Next, we use the HAM to obtain analytical approximate solutions for hairy black holes in Einstein-Weyl-AdS gravity. To achieve this, we perform a coordinate transformation $z=\frac{r_0}{r}$, such that the region of $r\rightarrow\infty$ becomes a finite value $z=0$. Then, substituting the potential function \eq{eqv} into Eqs.\eqref{eqphi} and \eqref{eqmu}, the field equations under this coordinate transformation become
\begin{eqnarray}
  &&(z-1) \Big[\Lambda  r_0^2+z^2 (\Lambda  r_0^2-3)+\Lambda  r_0^2 z \Big] \Big[(z-1) (\Lambda  r_0^2+z^2 (\Lambda  r_0^2-3)+\Lambda  r_0^2 z) \nonumber\\
  &&*(-3 \alpha  z^2 (-2 \Lambda  r_0^2+z^3 (\Lambda  r_0^2-3)+4 z^2) \mu ''(z)+4 z (-\Lambda  r_0^2+z^3 (\Lambda  r_0^2-3)+3 z^2){}^2 \phi '(z) \phi ''(z) \nonumber\\
  &&+(-4 \Lambda ^2 r_0^4+6 z^6 (\Lambda  r_0^2-3){}^2+26 z^5 (\Lambda  r_0^2-3)-2 \Lambda  r_0^2 z^3 (\Lambda  r_0^2-3)+4 \Lambda  r_0^2 z^2+24 z^4) \phi '(z)^2) \nonumber\\
  &&+6 \alpha  z \mu '(z) (-\Lambda ^2 r_0^4+z^6 (-(\Lambda  r_0^2-3){}^2)-6 z^5 (\Lambda  r_0^2-3)+\Lambda  r_0^2 z^3 (3-\Lambda  r_0^2)+3 \Lambda  r_0^2 z^2  \nonumber\\
  &&+(z^4 (\Lambda  r_0^2-3)-\Lambda  r_0^2 z+3 z^3){}^2 \phi '(z)^2-12 z^4) \Big]+6 \alpha  \mu (z) \Big[-4 \Lambda ^3 r_0^6+\Lambda ^3 r_0^6 z^9-9 \Lambda ^2 r_0^4 z^9  \nonumber\\
  &&+27 \Lambda  r_0^2 z^9+10 \Lambda ^2 r_0^4 z^8-60 \Lambda  r_0^2 z^8+36 \Lambda  r_0^2 z^7-20 \Lambda ^2 r_0^4 z^5+60 \Lambda  r_0^2 z^5-36 \Lambda  r_0^2 z^4+12 \Lambda ^3 r_0^6 z^3  \nonumber\\
  &&-36 \Lambda ^2 r_0^4 z^3+28 \Lambda ^2 r_0^4 z^2 +2 (z^4 (\Lambda  r_0^2-3)-\Lambda  r_0^2 z+3 z^3){}^3 \phi '(z) \phi ''(z)+2 z^4 (z (\Lambda  r_0^2-3)+2)  \nonumber\\
  &&*(-\Lambda  r_0^2+z^3 (\Lambda  r_0^2-3)+3 z^2){}^2 \phi '(z)^2-27 z^9+90 z^8-108 z^7+36 z^6 \Big]=0 \label{eqphiz},
\end{eqnarray}

\begin{eqnarray}
  &&3 \alpha ^2 \beta (9 (\Lambda  r_0^2-3){}^4 z^8+68 (\Lambda  r_0^2-3){}^3 z^7+164 (\Lambda  r_0^2-3){}^2 z^6-12 (3 \Lambda ^4 r_0^8-27 \Lambda ^3 r_0^6+81 \Lambda ^2 r_0^4     \nonumber\\
  &&-97 \Lambda  r_0^2+48) z^5-8 (17 \Lambda ^3 r_0^6-102 \Lambda ^2 r_0^4+153 \Lambda  r_0^2-18) z^4-208 \Lambda  r_0^2 (\Lambda  r_0^2-3) z^3  \nonumber\\
  &&-12 \Lambda  r_0^2 (3 \Lambda ^3 r_0^6-18 \Lambda ^2 r_0^4+27 \Lambda  r_0^2+16) z^2-112 \Lambda ^2 r_0^4 (\Lambda  r_0^2-3) z-64 \Lambda ^2 r_0^4) \mu (z)^2 z^6       \nonumber\\
  &&-(z-1) \alpha  ((\Lambda  r_0^2-3) z^2+\Lambda  r_0^2 z+\Lambda  r_0^2) \mu (z) (12 \alpha  \beta  \Lambda ^3 r_0^6 \mu ''(z) z^{13}-108 \alpha  \beta  \Lambda ^2 r_0^4 \mu ''(z) z^{13}   \nonumber\\
  &&+324 \alpha  \beta  \Lambda  r_0^2 \mu ''(z) z^{13}-324 \alpha  \beta  \mu ''(z) z^{13}-8 \beta  \Lambda ^4 r_0^8 z^{12}+96 \beta  \Lambda ^3 r_0^6 z^{12}-432 \beta  \Lambda ^2 r_0^4 z^{12}   \nonumber\\
  &&+864 \beta  \Lambda  r_0^2 z^{12}-648 \beta  z^{12}+84 \alpha  \beta  \Lambda ^2 r_0^4 \mu ''(z) z^{12}-504 \alpha  \beta  \Lambda  r_0^2 \mu ''(z) z^{12}+756 \alpha  \beta  \mu ''(z) z^{12}   \nonumber\\
  &&-64 \beta  \Lambda ^3 r_0^6 z^{11}+576 \beta  \Lambda ^2 r_0^4 z^{11}-1728 \beta  \Lambda  r_0^2 z^{11}+1728 \beta  z^{11}+192 \alpha  \beta  \Lambda  r_0^2 \mu ''(z) z^{11}   \nonumber\\
  &&-576 \alpha  \beta  \mu ''(z) z^{11}-176 \beta  \Lambda ^2 r_0^4 z^{10}+1056 \beta  \Lambda  r_0^2 z^{10}-1584 \beta  z^{10}-12 \alpha  \beta  \Lambda ^3 r_0^6 \mu ''(z) z^{10}   \nonumber\\
  &&+72 \alpha  \beta  \Lambda ^2 r_0^4 \mu ''(z) z^{10}-108 \alpha  \beta  \Lambda  r_0^2 \mu ''(z) z^{10}+144 \alpha  \beta  \mu ''(z) z^{10}+40 \beta  \Lambda ^4 r_0^8 z^9-3 \Lambda ^3 r_0^8 z^9   \nonumber\\
  &&-360 \beta  \Lambda ^3 r_0^6 z^9+27 \Lambda ^2 r_0^6 z^9+1080 \beta  \Lambda ^2 r_0^4 z^9-81 \Lambda  r_0^4 z^9-1368 \beta  \Lambda  r_0^2 z^9+81 r_0^2 z^9+864 \beta  z^9   \nonumber\\
  &&-48 \alpha  \beta  \Lambda ^2 r_0^4 \mu ''(z) z^9+144 \alpha  \beta  \Lambda  r_0^2 \mu ''(z) z^9+152 \beta  \Lambda ^3 r_0^6 z^8-9 \Lambda ^2 r_0^6 z^8-912 \beta  \Lambda ^2 r_0^4 z^8   \nonumber\\
  &&+54 \Lambda  r_0^4 z^8+1368 \beta  \Lambda  r_0^2 z^8-81 r_0^2 z^8-288 \beta  z^8-48 \alpha  \beta  \Lambda  r_0^2 \mu ''(z) z^8+256 \beta  \Lambda ^2 r_0^4 z^7   \nonumber\\
  &&+36 \Lambda  r_0^4 z^7-768 \beta  \Lambda  r_0^2 z^7-108 r_0^2 z^7+6 \alpha  \beta  (5 (\Lambda  r_0^2-3){}^3 z^5+36 (\Lambda  r_0^2-3){}^2 z^4  \nonumber\\
  &&+76 (\Lambda  r_0^2-3) z^3-2 (7 \Lambda ^3 r_0^6-42 \Lambda ^2 r_0^4+63 \Lambda  r_0^2-24) z^2-48 \Lambda  r_0^2 (\Lambda  r_0^2-3) z-40 \Lambda  r_0^2) \mu '(z) z^7    \nonumber\\
  &&+40 \beta  \Lambda ^4 r_0^8 z^6+3 \Lambda ^3 r_0^8 z^6-240 \beta  \Lambda ^3 r_0^6 z^6-18 \Lambda ^2 r_0^6 z^6+360 \beta  \Lambda ^2 r_0^4 z^6+27 \Lambda  r_0^4 z^6+288 \beta  \Lambda  r_0^2 z^6   \nonumber\\
  &&+108 r_0^2 z^6+128 \beta  \Lambda ^3 r_0^6 z^5-36 \Lambda ^2 r_0^6 z^5-384 \beta  \Lambda ^2 r_0^4 z^5+108 \Lambda  r_0^4 z^5+64 \beta  \Lambda ^2 r_0^4 z^4-144 \Lambda  r_0^4 z^4   \nonumber\\
  &&+12 \Lambda ^3 r_0^8 z^3-36 \Lambda ^2 r_0^6 z^3+72 \Lambda ^2 r_0^6 z^2-6 r_0^2 ((\Lambda  r_0^2-3) z^3+3 z^2-\Lambda  r_0^2){}^3 \phi '(z)^2 z^2-12 \Lambda ^3 r_0^8)   \nonumber\\
  &&+((\Lambda  r_0^2-3) z^3+3 z^2-\Lambda  r_0^2){}^2 \Big[3 \alpha ^2 \beta  (z (\Lambda  r_0^2-3)+2){}^2 \mu '(z)^2 z^8-z \alpha  (-24 (9 z^3-24 z^2+18 z     \nonumber\\
  &&-4) \beta  z^6+3 (72 \beta  \Lambda  z^5-128 \beta  \Lambda  z^4+48 \beta  \Lambda  z^3+(9-96 \beta  \Lambda ) z^2+7 (16 \beta  \Lambda -3) z-32 \beta  \Lambda +12) r_0^2 z^4     \nonumber\\
  &&+\Lambda  (-72 \beta  \Lambda  z^7+64 \beta  \Lambda  z^6+6 (32 \beta  \Lambda -3) z^4-7 (16 \beta  \Lambda -3) z^3+27 z-30) r_0^4 z^2     \nonumber\\
  &&+\Lambda ^2 (8 \beta  \Lambda  z^9+(3-32 \beta  \Lambda ) z^6-9 z^3+6) r_0^6) \mu '(z)+2 (z-1) ((\Lambda  r_0^2-3) z^2+\Lambda  r_0^2 z+\Lambda  r_0^2)    \nonumber\\
  &&*(r_0^2 ((\Lambda  r_0^2-3) z^3+3 z^2-\Lambda  r_0^2){}^2 \phi '(z)^2-2 z^6 \alpha  \beta  (z (\Lambda  r_0^2-3)+2){}^2 \mu ''(z))\Big]=0  \label{eqmuz},
\end{eqnarray}
where the prime ($'$) denotes the differentiation of the function with respect to $z$.
In the two above equations, we use the Schwarzschild AdS metric as our seed solution, which is
$h(r)=f(r)=1-\frac{2M}{r}-\frac{\Lambda}{3}r^2$,
and $M=\frac{r_0}{6}(3-\Lambda r_0^2)$, where $r_0$ represents the event horizon of the Schwarzschild AdS black hole.
Note that both Eqs.\eqref{eqphiz} and \eqref{eqmuz} are second-order derivative equations with respect to $\phi(z)$ and $\mu(z)$. Then, we apply the HAM to the two nonlinear equations Eqs.\eqref{eqphiz} and \eqref{eqmuz} to obtain the analytical approximate solutions of $\phi(z)$ and $\mu(z)$.

In the homotopy equation, the auxiliary function can be coded into the initial guess solution, that is, $H_i(z)$ on the right side of the zero-order deformation equation (Eq.~\eqref{hm}) can be moved to the left side, as described in Ref.\cite{Robert2009}. Therefore, without loss of generality, we set the auxiliary function $H_i(z)=1, i=1,2$. The initial approximations are taken as
\begin{eqnarray}
  &&\phi_0(z)=Q*z \label{phi0}\\
  &&\mu_0(z)=z(1-z) \label{mu0}
\end{eqnarray}

where the constant $Q$ represents the scalar charge on the event horizon $z=1 ~(r=r_0)$.

Because both differential equations are second
order, the auxiliary linear operators (whose construction method is shown in Ref.\cite{Robert2009,Yinshan2010}) are given by
\begin{eqnarray}
  L\left[\phi_i(z;q)\right]=\frac{\partial^2\phi_i(z;q)}{\partial z^2},\quad i=1,2.
\end{eqnarray}

The following boundary conditions are employed during the solution process of the nonlinear equations using the HAM:
\begin{eqnarray}
  \phi(0)=0,~\phi(1)=Q; \\
  \mu(0)=0=\mu(1).
\end{eqnarray}
The selected initial approximations Eqs.\eqref{phi0} and \eqref{mu0} are required to satisfy the boundary conditions for the system of differential equations above.

The nonlinear operators $N_i$ are provided by Eqs.\eqref{eqphiz} and \eqref{eqmuz}. The initial approximations Eqs.\eqref{phi0} and \eqref{mu0} are substituted into the $m$th-order deformation equation \eq{m-orderhm}. Then,
by solving each $y_{im}(z)$ of \eq{m-orderhm} with the boundary conditions, we can obtain the $M$-order analytical approximate solution ($y_i^M(z)$) \eq{m-y-expand}.
Here, we perform the fourth-order approximation ($M=4$) for both functions. Therefore, letting $Q=1,\alpha=1,~ \beta=1/2, ~\Lambda=-6/100, ~r_0=1$ and fixing the convergence control parameters $h_1=h_2=h$, the solutions are related to $z$ and $h$, which are shown in Eqs.\eqref{solphiz} and \eqref{solmuz} of the appendix.

To select an optimal value of $h$, we substitute Eqs. \eqref{solphiz} and \eqref{solmuz} into the left side of the nonlinear equations \eqref{eqphiz} and \eqref{eqmuz} and define
\begin{eqnarray}
\label{delta eq1}
&&\Delta eq_1=|N_1[\phi(z),\mu(z)]|,\\
\label{delta eq2}
&&\Delta eq_2=|N_2[\phi(z),\mu(z)]|,
\end{eqnarray}
which represent the deviations between the analytical approximate solutions and the exact solutions. Obviously, when the two functions $\Delta eq_i ~(i=1,2)$ are as close to zero as possible in $z\in[0,1]$, the obtained approximate solutions are as close to the analytical solutions as possible. Here, we use the averaged square residual error function (detailed in Ref.\cite{Liao:2003mua}) to determine the optimal value $h$, which represents the total deviation between the approximate and exact solutions
\begin{eqnarray}
E(h)=\frac{1}{S+1}\sum_{k=0}^{S}\Bigg\{\Big(N_1[\phi(z_k,h),
\mu(z_k,h)]\Big)^2+\Big(N_2[\phi(z_k,h),\mu(z_k,h)]\Big)^2 \Bigg\},
\label{totalerror}
\end{eqnarray}
with
\begin{eqnarray}
z_k=k\Delta z=k\frac{1}{S},~~ k=0,1,2,...,S.
\label{zk}
\end{eqnarray}
Hereafter, a value of $S=40$ is used for the purpose of optimization. Then, we plot the logarithm of the averaged square residual error function $\ln{E(h)}$ as the undetermined parameter $h$ changes.

\begin{figure}[htb]
\centering
\label{figlogem} 
\subfigure[]{
\label{figlogem1} 
\includegraphics[width=2.9in]{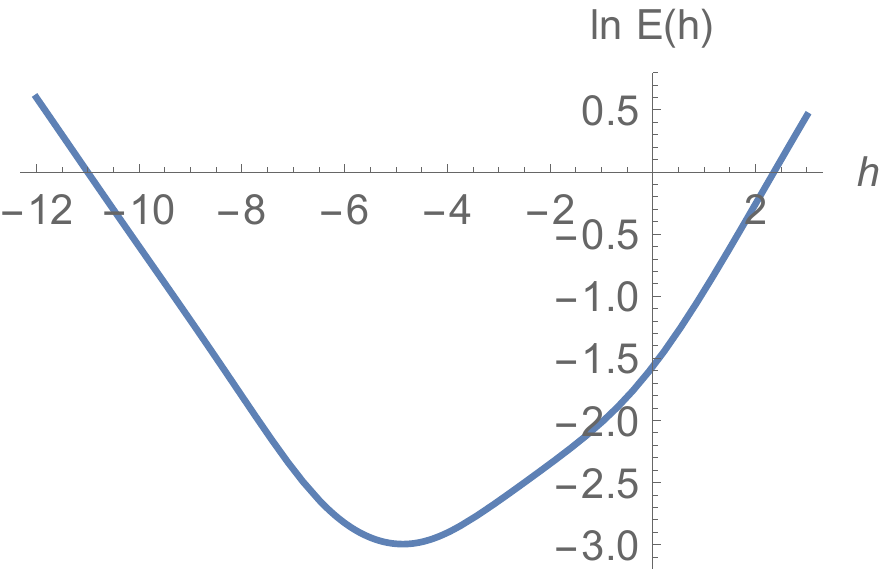}}
\quad
\subfigure[]{
\label{figlogem2} 
\includegraphics[width=2.9in]{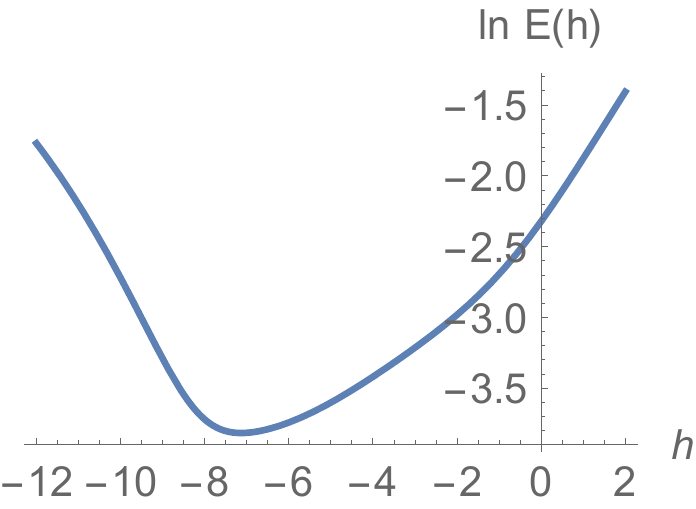}}
\quad
\subfigure[]{
\label{figlogem3} 
\includegraphics[width=2.9in]{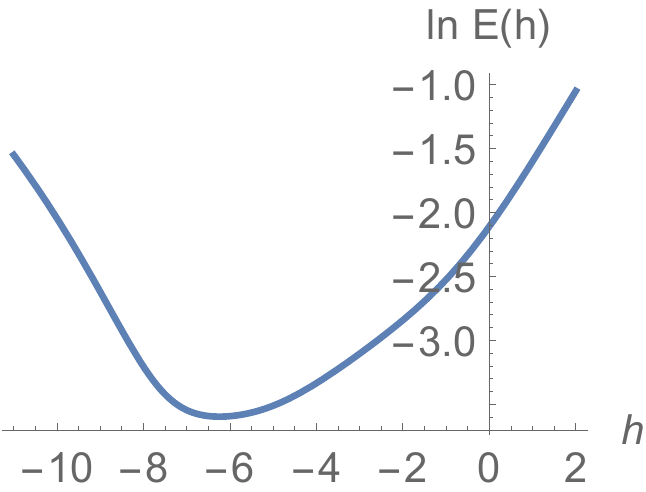}}
\quad
\subfigure[]{
\label{figlogem4} 
\includegraphics[width=2.9in]{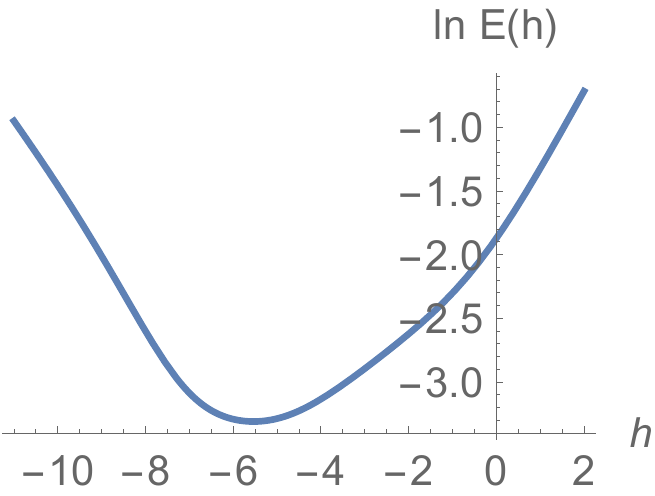}}
\quad
\caption{Plot of the logarithm of the square residual $ln\left[E(h)\right]$ as a function of $h$ for different values of $\alpha$ with fixed $Q=1,~\alpha=1,~ \beta=1/2, ~\Lambda=-6/100$, and $r_0=1$.}\label{figem}
\end{figure}

Considering different values of $\alpha$, Fig.\ref{figem} shows that the logarithm of the square residual $ln\left[E(h)\right]$ as a function of $h$. The lowest point of the curve represents the minimum value of the averaged square residual error. Therefore, we can mathematically determine the optimal value of $h$ by setting
\begin{eqnarray}
h^*=min\{E(h)\}.
\label{opth}
\end{eqnarray}

Substituting Eqs.\ref{solphiz} and \ref{solmuz} into Eqs.\eqref{totalerror} and \eqref{opth}, the optimal value of $h$ can be identified. We obtain $h^*=-4.866$ for $\alpha=1$, $h^*=-7.131$ for $\alpha=1/2$, $h^*=-6.242$ for $\alpha=2/3$, and $h^*=-5.536$ for $\alpha=4/5$. After reverting back to the radial coordinate $r$, the analytical approximate solutions of Einstein-Weyl-AdS-scalar gravity are determined.

Taking $Q=1,~\alpha=1,~ \beta=1/2, ~\Lambda=-6/100,~ and ~r_0=1$, we have $h=-4.866$, and the averaged square residual error has the minimum value $E(h)=0.001011$. Then, the analytical approximate solutions are obtained as
\begin{eqnarray}
  \phi(r)=&&\frac{1}{r^{69}}\Big(1.39206 r^{68}+0.000745152 r^{67}-0.0119691 r^{66}+0.0144891 r^{65}-0.44635 r^{64}+0.615164 r^{63} \nonumber\\
  &&-2.72805 r^{62}-3.92652 r^{61}+12.6285 r^{60}-137.351 r^{59}+366.534 r^{58}-216.3 r^{57}-226.609 r^{56} \nonumber\\
  &&+471.467 r^{55}+368.19 r^{54}-3813.81 r^{53}+26514.1 r^{52}-116346. r^{51}+379609. r^{50} \nonumber\\
  &&-1.07828\times 10^6 r^{49}+2.11948\times 10^6 r^{48}-2.29864\times 10^6 r^{47}+1.02257\times 10^6 r^{46} \nonumber\\
  &&-1.50333\times 10^6 r^{45}+1.18351\times 10^7 r^{44}-5.41334\times 10^7 r^{43}+2.04949\times 10^8 r^{42} \nonumber\\
  &&-6.48124\times 10^8 r^{41}+1.71142\times 10^9 r^{40}-3.73141\times 10^9 r^{39}+6.11124\times 10^9 r^{38} \nonumber\\
  &&-6.49917\times 10^9 r^{37}+4.46297\times 10^9 r^{36}-9.7344\times 10^9 r^{35}+4.79214\times 10^{10} r^{34} \nonumber\\
  &&-1.73721\times 10^{11} r^{33}+5.18957\times 10^{11} r^{32}-1.34733\times 10^{12} r^{31}+2.99473\times 10^{12} r^{30} \nonumber\\
  &&-5.46302\times 10^{12} r^{29}+7.45301\times 10^{12} r^{28}-5.96958\times 10^{12} r^{27}-5.30202\times 10^{11} r^{26} \nonumber\\
 &&+6.88219\times 10^{12} r^{25}-4.01186\times 10^{12} r^{24}-7.91257\times 10^{12} r^{23}+1.57059\times 10^{13} r^{22} \nonumber\\
  &&-1.02945\times 10^{13} r^{21}-8.17116\times 10^{11} r^{20}+7.00387\times 10^{12} r^{19}-9.69189\times 10^{12} r^{18} \nonumber\\
  &&+1.4822\times 10^{13} r^{17}-2.10735\times 10^{13} r^{16}+2.32064\times 10^{13} r^{15}-1.80509\times 10^{13} r^{14} \nonumber\\
  &&+4.34315\times 10^{12} r^{13}+1.43146\times 10^{13} r^{12}-2.67435\times 10^{13} r^{11}+2.36935\times 10^{13} r^{10} \nonumber\\
  &&-9.32602\times 10^{12} r^9-3.52587\times 10^{12} r^8+7.54084\times 10^{12} r^7-5.2478\times 10^{12} r^6 \nonumber\\
  &&+2.07873\times 10^{12} r^5-4.51641\times 10^{11} r^4+2.26116\times 10^{10} r^3+1.3875\times 10^{10} r^2 \nonumber\\
  &&-3.34599\times 10^9 r+2.40752\times 10^8 \Big) \label{solphir}
\end{eqnarray}
\begin{eqnarray}
  \mu(r)=&&\frac{1}{r^{74}}\Big( 0.873958 r^{73}-1.00002 r^{72}+0.000718144 r^{71}-0.00143588 r^{70}+0.0351733 r^{69}  \nonumber\\
  &&-0.0891202 r^{68}+1.06933 r^{67}-3.30994 r^{66}+19.7923 r^{65}-66.3907 r^{64}+209.652 r^{63}  \nonumber\\
  &&-602.712 r^{62}+1053.06 r^{61}-1095.22 r^{60}+870.997 r^{59}-497.753 r^{58}+120.428 r^{57}  \nonumber\\
  &&-1396.33 r^{56}+11019.1 r^{55}-48659.1 r^{54}+160275. r^{53}-462283. r^{52}+1.0942\times 10^6 r^{51}  \nonumber\\
  &&-2.06221\times 10^6 r^{50}+3.48541\times 10^6 r^{49}-5.87153\times 10^6 r^{48}+9.74318\times 10^6 r^{47}  \nonumber\\
  &&-1.78659\times 10^7 r^{46}+4.44373\times 10^7 r^{45}-1.26768\times 10^8 r^{44}+3.33937\times 10^8 r^{43}  \nonumber\\
  &&-7.61599\times 10^8 r^{42}+1.43689\times 10^9 r^{41}-2.19665\times 10^9 r^{40}+3.13804\times 10^9 r^{39}  \nonumber\\
  &&-5.93324\times 10^9 r^{38}+1.49292\times 10^{10} r^{37}-3.84801\times 10^{10} r^{36}+9.43637\times 10^{10} r^{35}  \nonumber\\
  &&-2.17726\times 10^{11} r^{34}+4.58949\times 10^{11} r^{33}-8.52325\times 10^{11} r^{32}+1.33135\times 10^{12} r^{31}  \nonumber\\
  &&-1.65726\times 10^{12} r^{30}+1.64475\times 10^{12} r^{29}-1.71411\times 10^{12} r^{28}+2.86199\times 10^{12} r^{27}  \nonumber\\
  &&-5.52305\times 10^{12} r^{26}+9.20825\times 10^{12} r^{25}-1.43825\times 10^{13} r^{24}+2.31343\times 10^{13} r^{23}  \nonumber\\
  &&-3.51017\times 10^{13} r^{22}+4.28285\times 10^{13} r^{21}-3.28237\times 10^{13} r^{20}-7.70712\times 10^{12} r^{19}  \nonumber\\
  &&+7.98116\times 10^{13} r^{18}-1.59981\times 10^{14} r^{17}+1.99413\times 10^{14} r^{16}-1.49078\times 10^{14} r^{15}  \nonumber\\
  &&+1.23135\times 10^{12} r^{14}+1.81712\times 10^{14} r^{13}-2.85882\times 10^{14} r^{12}+2.28652\times 10^{14} r^{11}  \nonumber\\
  &&-4.72387\times 10^{13} r^{10}-1.19066\times 10^{14} r^9+1.60953\times 10^{14} r^8-9.44017\times 10^{13} r^7  \nonumber\\
  &&+1.17265\times 10^{13} r^6+2.62055\times 10^{13} r^5-2.41191\times 10^{13} r^4+1.11424\times 10^{13} r^3  \nonumber\\
  &&-3.09446\times 10^{12} r^2+4.94441\times 10^{11} r-3.52851\times 10^{10} \Big) \label{solmur}
\end{eqnarray}

\begin{figure}[htb]
\centering
\subfigure[]{
\label{phir} 
\includegraphics[width=2.9in]{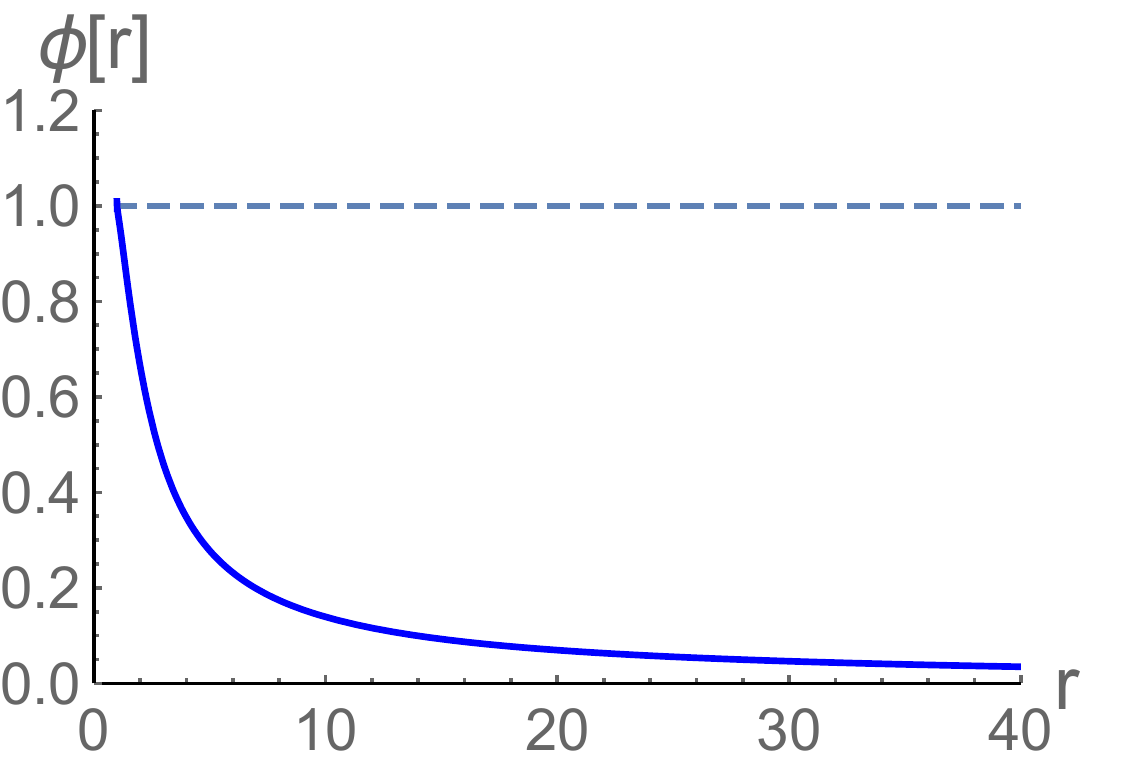}}
\quad
\subfigure[]{
\label{mur} 
\includegraphics[width=2.9in]{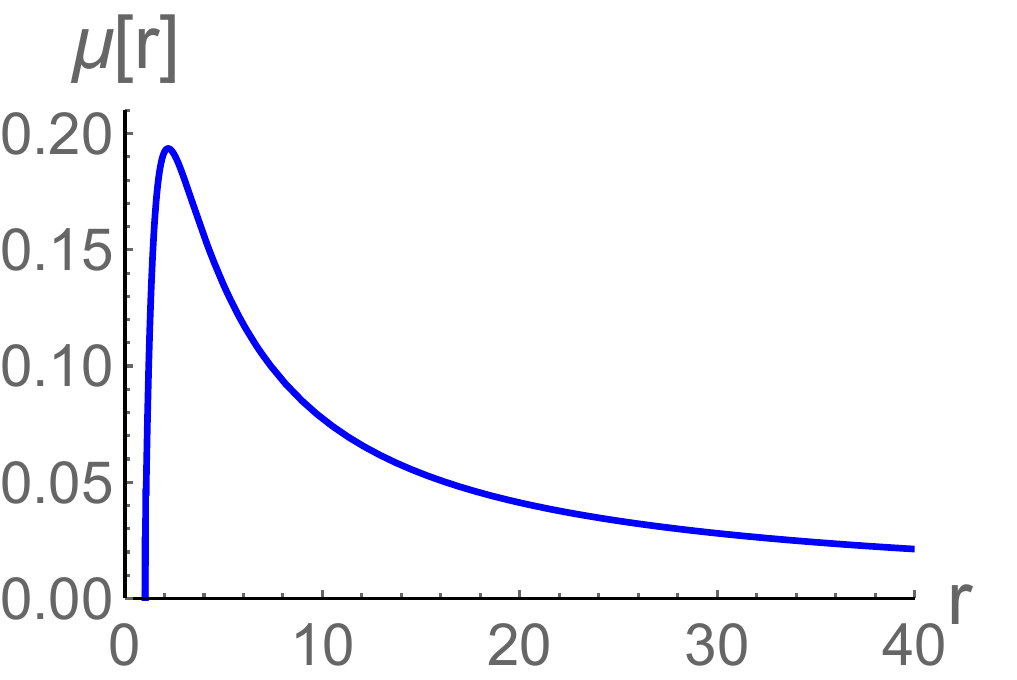}}
\caption{The scalar field $\phi(r)$ and the deformation function $\mu(r)$ with $Q=1,~\alpha=1,~ \beta=1/2, ~\Lambda=-6/100, ~r_0=1$, and $h^*=-4.866$.}
\label{phimu}
\end{figure}

\begin{figure}[htb]
\centering
\includegraphics[width=3.2in]{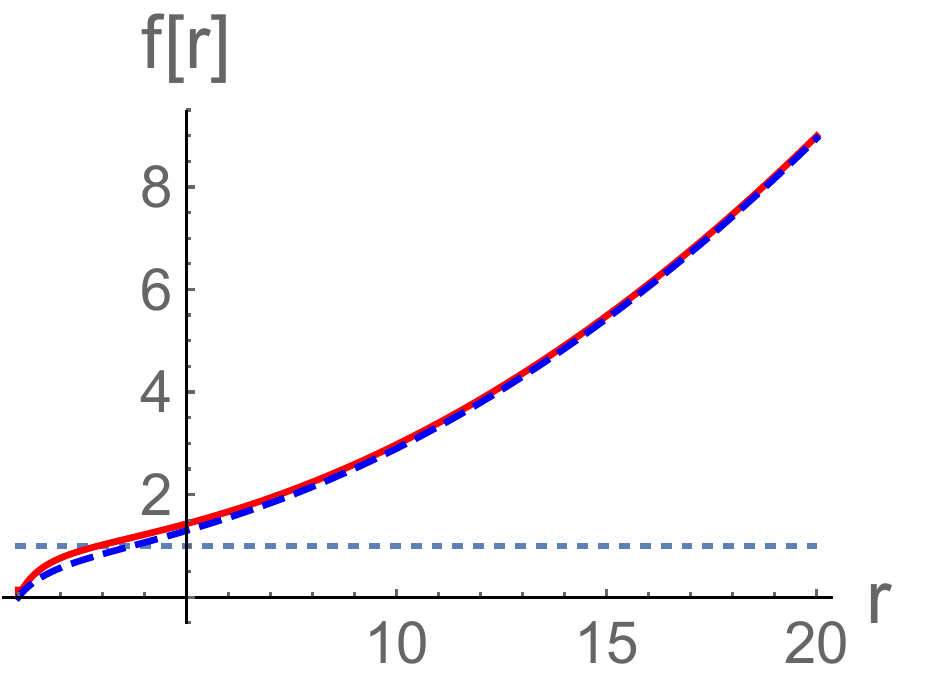}
\caption{Metric function $(g_{rr})^{-1}$ for the Schwarzschild-AdS metric (blue dashed line), and the generated solution (red solid line) with $Q=1,~\alpha=1,~ \beta=1/2, ~\Lambda=-6/100, ~r_0=1$, and $h^*=-4.866$.}\label{fr}
\end{figure}

The scalar field and deformation function are plotted in Fig.\ref{phimu}, where we can see that $\phi(r_0)=1, \mu(r_0)=0$. Both functions disappear for large $r$; hence, the behavior of the generated solution is similar to the seed Schwarzschild-AdS metric. The metric function $g_{rr}=f(r)+\alpha \mu(r)$ for the generated metric is shown
in Fig.\ref{fr}. For comparison, we also include the corresponding Schwarzschild-AdS metric function $f(r)=1-\frac{2M}{r}-\frac{\Lambda}{3}r^2$ as a blue dashed line. Because the metric function $h(r)=1-\frac{2M}{r}-\frac{\Lambda}{3}r^2$ is unchanged in the MGD method, the generated solution has the same metric function $h(r)$ with the seed Schwarzschild-AdS metric.

\begin{figure}[htb]
\centering
\includegraphics[width=3.2in]{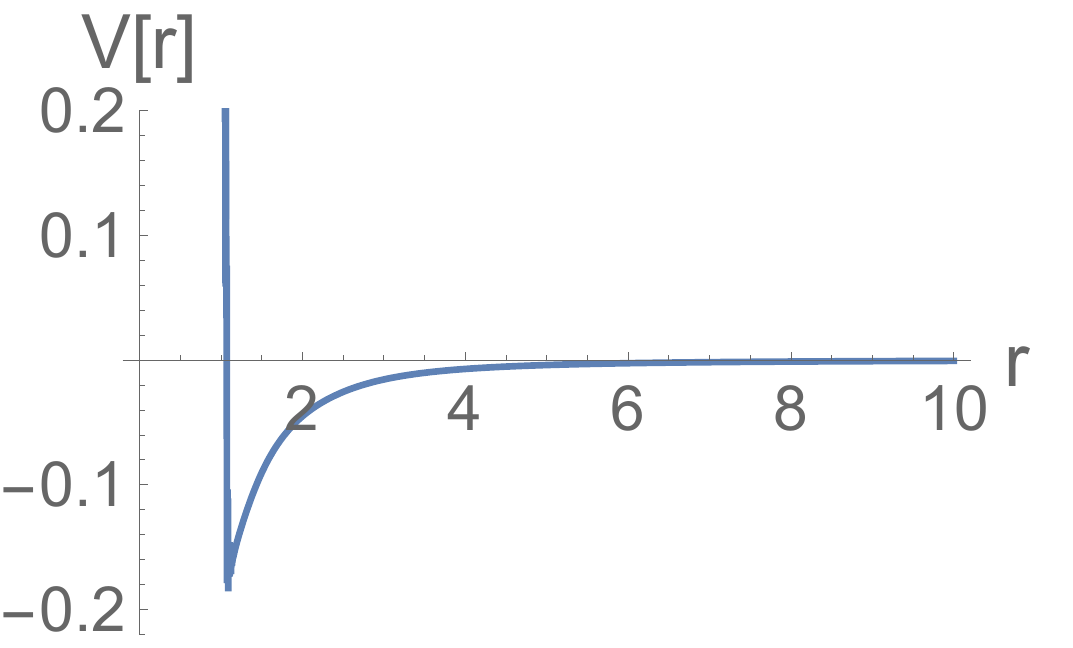}
\caption{Potential function $V(r)$ with $Q=1,~\alpha=1,~ \beta=1/2, ~\Lambda=-6/100, ~r_0=1$, and $h^*=-4.866$.}\label{vr}
\end{figure}

The potential function $V(r)$ can be determined by substituting $\phi(r)$ of Eq. \eqref{solphir} and $\mu(r)$ Eq. \eqref{solmur} into Eq. \eqref{eqv}, which is depicted in Fig.\ref{vr}. We find that the potential disappears at infinity as expected.

\begin{figure}[htb]
\centering
\label{err}
\subfigure[]{
\label{err1} 
\includegraphics[width=2.9in]{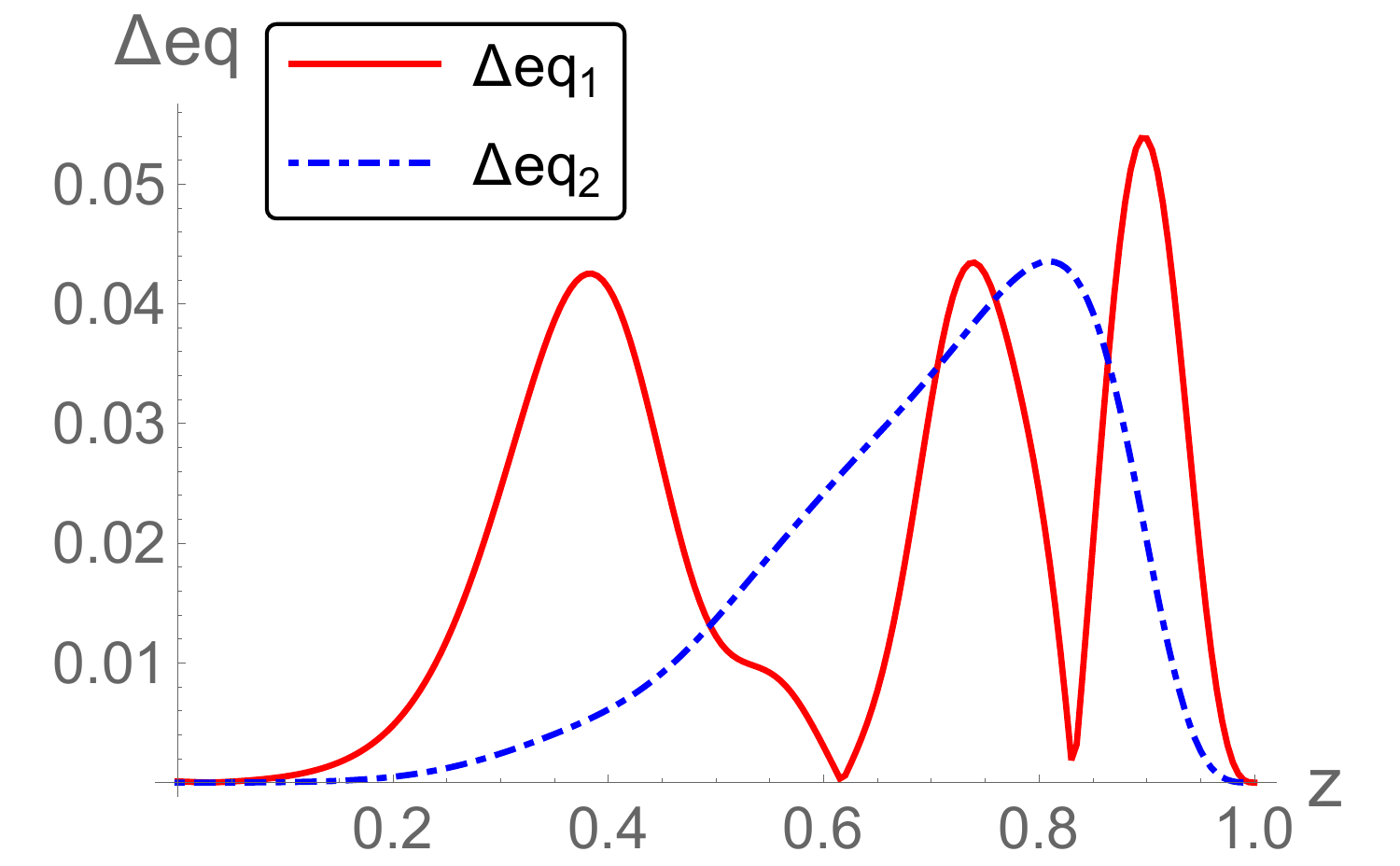}}
\quad
\subfigure[]{
\label{err2} 
\includegraphics[width=2.9in]{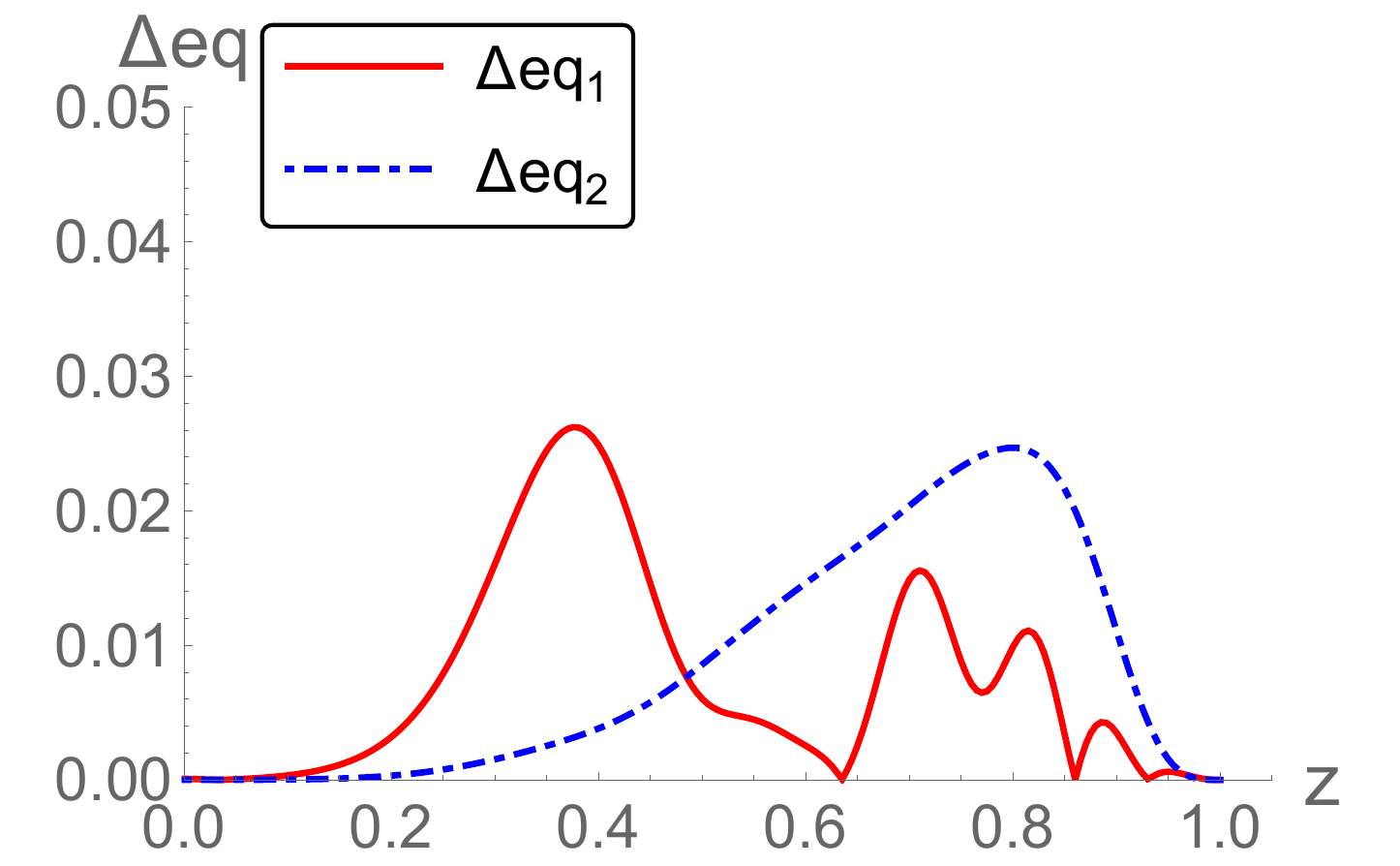}}
\caption{Absolute errors of the analytic approximate solutions from the two field equations, taking $Q=1,~\alpha=1,~ \beta=1/2, ~\Lambda=-6/100, ~r_0=1$, which represent the deviations of the analytical approximate
solutions from the exact solutions.}
\end{figure}

It is also interesting to check the accuracy of the analytical approximate solutions. Taking different values of $\alpha$, two curves in Fig.\ref{err} show the deviations (Eqs. \eqref{delta eq1} and \eqref{delta eq2}) of the analytical approximate
solutions from the exact solutions in the entire spacetime outside the event horizon ($z\in[0,1]$). We find that analytical approximate solutions with larger values of $\alpha$ are more accurate than those with smaller $\alpha$.

\section{Conclusions and discussions}
\label{conc}

In this study, we investigate Einstein-Weyl gravity with a minimally coupled scalar field in four dimensional spacetime. From the corresponding field equations, we introduce geometric deformation to the radial metric component using the MGD approach, which splits the highly nonlinear and coupled field equations into two subsystems. One subsystem describes the background geometry, whereas the other incorporates the scalar field source. Taking the Schwarzschild-AdS metric as the seed solution, we substitute this into the deformed equations to obtain a simplified system for the deformation and scalar field. The HAM was then utilized to derive fourth order analytical approximations satisfying appropriate boundary conditions. An optimal convergence control parameter is determined mathematically by minimizing the averaged residual error.
We discuss the accuracy of these approximations over the entire exterior spacetime.

The results demonstrate the efficacy of combining the MGD and HAM techniques for generating new solutions in modified gravity theories. The analytical approximations provide valuable physical insights and can be used to study the thermodynamic and dynamical properties of these black holes.
This approach, which can obtain hairy black hole solutions in higher-order gravity theories, provides a useful addition and extension to numerical methods to find such solutions. The demonstrated accuracy and flexibility of the method offer promise for the handling of more complex scenarios and exploring other gravitational models.


\section*{APPENDIX}
\label{APPENDIX}
Choosing $Q=1$, $\alpha=1$, $\beta=1/2$, $\Lambda=-6/100$, $r_0=1$ and the convergence control parameters  $h_1=h_2=h$, the solutions are related to z and h, and the fourth-order approximation $M=4$ solutions can be written as

\begin{eqnarray}
  \phi(z,h)=&&429410. h^4 z^{69}-5.96797\times 10^6 h^4 z^{68}+2.47478\times 10^7 h^4 z^{67}+4.03304\times 10^7 h^4 z^{66}  \nonumber\\
  &&-8.05555\times 10^8 h^4 z^{65}+3.70766\times 10^9 h^4 z^{64}-9.36008\times 10^9 h^4 z^{63}+1.345\times 10^{10} h^4 z^{62}  \nonumber\\
  &&-6.2888\times 10^9 h^4 z^{61}-1.66341\times 10^{10} h^4 z^{60}+4.22602\times 10^{10} h^4 z^{59}-4.77001\times 10^{10} h^4 z^{58}  \nonumber\\
  &&+2.55318\times 10^{10} h^4 z^{57}+7.74652\times 10^9 h^4 z^{56}-3.21959\times 10^{10} h^4 z^{55}+4.13913\times 10^{10} h^4 z^{54}  \nonumber\\
  &&-3.75871\times 10^{10} h^4 z^{53}+2.64368\times 10^{10} h^4 z^{52}-1.72866\times 10^{10} h^4 z^{51}+8642.82 h^3 z^{51}  \nonumber\\
  &&+1.24922\times 10^{10} h^4 z^{50}-92007. h^3 z^{50}-1.45737\times 10^9 h^4 z^{49}+252862. h^3 z^{49}  \nonumber\\
  &&-1.83613\times 10^{10} h^4 z^{48}+779721. h^3 z^{48}+2.80118\times 10^{10} h^4 z^{47}-7.24849\times 10^6 h^3 z^{47}  \nonumber\\
  &&-1.41084\times 10^{10} h^4 z^{46}+2.22141\times 10^7 h^3 z^{46}-7.16285\times 10^9 h^4 z^{45}-3.51403\times 10^7 h^3 z^{45}  \nonumber\\
  &&+1.22797\times 10^{10} h^4 z^{44}+2.21412\times 10^7 h^3 z^{44}-9.40884\times 10^8 h^4 z^{43}+2.33255\times 10^7 h^3 z^{43}  \nonumber\\
  &&-1.06619\times 10^{10} h^4 z^{42}-7.0111\times 10^7 h^3 z^{42}+1.33109\times 10^{10} h^4 z^{41}+8.55189\times 10^7 h^3 z^{41}  \nonumber\\
  &&-9.75874\times 10^9 h^4 z^{40}-7.20154\times 10^7 h^3 z^{40}+5.34953\times 10^9 h^4 z^{39}+3.92437\times 10^7 h^3 z^{39}  \nonumber\\
  &&-2.39996\times 10^9 h^4 z^{38}+1.54129\times 10^7 h^3 z^{38}+9.12389\times 10^8 h^4 z^{37}-6.43894\times 10^7 h^3 z^{37}  \nonumber\\
  &&-2.9778\times 10^8 h^4 z^{36}+5.87428\times 10^7 h^3 z^{36}+8.30197\times 10^7 h^4 z^{35}-1.19406\times 10^7 h^3 z^{35}  \nonumber\\
  &&-1.88269\times 10^7 h^4 z^{34}-7.12623\times 10^6 h^3 z^{34}+3.64151\times 10^6 h^4 z^{33}-2.1015\times 10^7 h^3 z^{33}  \nonumber\\
  &&+111.56 h^2 z^{33}-1.34855\times 10^6 h^4 z^{32}+4.9845\times 10^7 h^3 z^{32}-826.039 h^2 z^{32}+999632. h^4 z^{31}  \nonumber\\
  &&-4.81757\times 10^7 h^3 z^{31}+1308.09 h^2 z^{31}-611717. h^4 z^{30}+2.94099\times 10^7 h^3 z^{30}  \nonumber\\
  &&+5286.38 h^2 z^{30}+284748. h^4 z^{29}-1.34729\times 10^7 h^3 z^{29}-23893.7 h^2 z^{29}-109144. h^4 z^{28}  \nonumber\\
  &&+5.1009\times 10^6 h^3 z^{28}+33274.2 h^2 z^{28}+34695.7 h^4 z^{27}-1.61135\times 10^6 h^3 z^{27}  \nonumber\\
  &&-6810.16 h^2 z^{27}-9054.92 h^4 z^{26}+420846. h^3 z^{26}-23958.5 h^2 z^{26}+2081.04 h^4 z^{25}  \nonumber\\
  &&-90010.7 h^3 z^{25}+12562.7 h^2 z^{25}-433.881 h^4 z^{24}+10791.4 h^3 z^{24}-705.319 h^2 z^{24}  \nonumber\\
  &&+79.9335 h^4 z^{23}+333.978 h^3 z^{23}+42918.5 h^2 z^{23}+24.8109 h^4 z^{22}+1707.58 h^3 z^{22}  \nonumber\\
  &&-89356.7 h^2 z^{22}-43.5262 h^4 z^{21}-2275.4 h^3 z^{21}+79470.4 h^2 z^{21}+23.4605 h^4 z^{20}  \nonumber\\
  &&+1234.53 h^3 z^{20}-40087. h^2 z^{20}-8.13607 h^4 z^{19}-420.612 h^3 z^{19}+14177.9 h^2 z^{19}  \nonumber\\
  &&+2.34391 h^4 z^{18}+126.822 h^3 z^{18}-4352.02 h^2 z^{18}-0.463647 h^4 z^{17}-26.4722 h^3 z^{17}  \nonumber\\
  &&+1001.93 h^2 z^{17}+0.0265813 h^4 z^{16}+2.93188 h^3 z^{16}-147.431 h^2 z^{16}+0.00841268 h^4 z^{15}  \nonumber\\
  &&-0.151642 h^3 z^{15}+17.3044 h^2 z^{15}+13.0983 h z^{15}-0.0038958 h^4 z^{14}+0.0389052 h^3 z^{14}  \nonumber\\
  &&+11.612 h^2 z^{14}-41.7552 h z^{14}-0.00374658 h^4 z^{13}-0.220038 h^3 z^{13}-5.03632 h^2 z^{13}  \nonumber\\
  &&+26.8412 h z^{13}-0.00104267 h^4 z^{12}-0.0742851 h^3 z^{12}-0.991202 h^2 z^{12}+41.2667 h z^{12}  \nonumber\\
  &&+0.00300242 h^4 z^{11}+0.182205 h^3 z^{11}+3.53299 h^2 z^{11}-62.1018 h z^{11}-0.00103572 h^4 z^{10}  \nonumber\\
  &&-0.0635955 h^3 z^{10}-1.19629 h^2 z^{10}+23.7919 h z^{10}+0.000097829 h^4 z^9+0.00640264 h^3 z^9  \nonumber\\
  &&+0.136406 h^2 z^9-2.07182 h z^9-0.0000333836 h^4 z^8-0.00292622 h^3 z^8-0.0657469 h^2 z^8  \nonumber\\
  &&+0.552441 h z^8-0.0000132256 h^4 z^7-0.000224176 h^3 z^7+0.00295546 h^2 z^7+0.578798 h z^7  \nonumber\\
  &&+4.30107\times 10^{-6} h^4 z^6+0.000139766 h^3 z^6+0.00120322 h^2 z^6-0.123379 h z^6  \nonumber\\
  &&-1.95624\times 10^{-6} h^4 z^5-0.000018467 h^3 z^5+0.000853395 h^2 z^5+0.0960924 h z^5  \nonumber\\
  &&+2.61877\times 10^{-7} h^4 z^4+0.0000146481 h^3 z^4+0.000262692 h^2 z^4-0.002016 h z^4  \nonumber\\
  &&-4.94209\times 10^{-8} h^4 z^3-2.90316\times 10^{-7} h^3 z^3+0.0000269415 h^2 z^3+0.002592 h z^3  \nonumber\\
  &&+1.3468\times 10^{-8} h^4 z^2+7.53332\times 10^{-7} h^3 z^2+0.0000135099 h^2 z^2-0.00010368 h z^2  \nonumber\\
  &&-0.000025954 h^4 z-0.00183426 h^3 z-0.0274575 h^2 z-0.173738 h z+z  \label{solphiz}\\
  \mu(z,h)=&&-6.29352\times 10^7 h^4 z^{74}+8.81894\times 10^8 h^4 z^{73}-5.51933\times 10^9 h^4 z^{72}+1.98737\times 10^{10} h^4 z^{71}  \nonumber\\
  &&-4.30193\times 10^{10} h^4 z^{70}+4.67406\times 10^{10} h^4 z^{69}+2.09156\times 10^{10} h^4 z^{68}-1.68377\times 10^{11} h^4 z^{67}  \nonumber\\
  &&+2.87079\times 10^{11} h^4 z^{66}-2.12369\times 10^{11} h^4 z^{65}-8.42558\times 10^{10} h^4 z^{64}+4.07827\times 10^{11} h^4 z^{63}  \nonumber\\
  &&-5.09905\times 10^{11} h^4 z^{62}+3.24105\times 10^{11} h^4 z^{61}+2.19626\times 10^9 h^4 z^{60}-2.65898\times 10^{11} h^4 z^{59}  \nonumber\\
  &&+3.55676\times 10^{11} h^4 z^{58}-2.85345\times 10^{11} h^4 z^{57}+1.42353\times 10^{11} h^4 z^{56}-808252. h^3 z^{56}  \nonumber\\
  &&-1.37448\times 10^{10} h^4 z^{55}+8.68926\times 10^6 h^3 z^{55}-5.85532\times 10^{10} h^4 z^{54}-4.01824\times 10^7 h^3 z^{54}  \nonumber\\
  &&+7.64102\times 10^{10} h^4 z^{53}+9.98154\times 10^7 h^3 z^{53}-6.26333\times 10^{10} h^4 z^{52}-1.23228\times 10^8 h^3 z^{52}  \nonumber\\
  &&+4.12606\times 10^{10} h^4 z^{51}-1.10655\times 10^7 h^3 z^{51}-2.55896\times 10^{10} h^4 z^{50}+3.08246\times 10^8 h^3 z^{50}  \nonumber\\
  &&+1.63154\times 10^{10} h^4 z^{49}-5.28697\times 10^8 h^3 z^{49}-9.7685\times 10^9 h^4 z^{48}+4.01497\times 10^8 h^3 z^{48}  \nonumber\\
  &&+5.11175\times 10^9 h^4 z^{47}+3.42747\times 10^7 h^3 z^{47}-3.15194\times 10^9 h^4 z^{46}-4.60404\times 10^8 h^3 z^{46}  \nonumber\\
  &&+3.06129\times 10^9 h^4 z^{45}+6.21299\times 10^8 h^3 z^{45}-3.05832\times 10^9 h^4 z^{44}-4.98284\times 10^8 h^3 z^{44}  \nonumber\\
  &&+2.42041\times 10^9 h^4 z^{43}+2.22845\times 10^8 h^3 z^{43}-1.51102\times 10^9 h^4 z^{42}+4.47977\times 10^7 h^3 z^{42}  \nonumber\\
  &&+7.7788\times 10^8 h^4 z^{41}-1.98092\times 10^8 h^3 z^{41}-3.41553\times 10^8 h^4 z^{40}+2.27667\times 10^8 h^3 z^{40}  \nonumber\\
  &&+1.30126\times 10^8 h^4 z^{39}-1.85797\times 10^8 h^3 z^{39}-4.34428\times 10^7 h^4 z^{38}+1.22579\times 10^8 h^3 z^{38}  \nonumber\\
  &&-4807.41 h^2 z^{38}+1.26814\times 10^7 h^4 z^{37}-6.78578\times 10^7 h^3 z^{37}+36009.6 h^2 z^{37}  \nonumber\\
  &&-3.22284\times 10^6 h^4 z^{36}+3.57908\times 10^7 h^3 z^{36}-107667. h^2 z^{36}+844519. h^4 z^{35}  \nonumber\\
  &&-2.30964\times 10^7 h^3 z^{35}+144354. h^2 z^{35}-366114. h^4 z^{34}+1.72808\times 10^7 h^3 z^{34}-12858.2 h^2 z^{34}  \nonumber\\
  &&+222798. h^4 z^{33}-1.14385\times 10^7 h^3 z^{33}-251171. h^2 z^{33}-121992. h^4 z^{32}+6.09981\times 10^6 h^3 z^{32}  \nonumber\\
  &&+405863. h^2 z^{32}+55607.2 h^4 z^{31}-2.69317\times 10^6 h^3 z^{31}-318568. h^2 z^{31}-21581.3 h^4 z^{30}  \nonumber\\
  &&+1.01577\times 10^6 h^3 z^{30}+99966.7 h^2 z^{30}+7062.34 h^4 z^{29}-328266. h^3 z^{29}+112145. h^2 z^{29}  \nonumber\\
  &&-1913.72 h^4 z^{28}+93668.6 h^3 z^{28}-253421. h^2 z^{28}+415.958 h^4 z^{27}-24418.9 h^3 z^{27}  \nonumber\\
  &&+282811. h^2 z^{27}-58.3501 h^4 z^{26}+5534.84 h^3 z^{26}-219658. h^2 z^{26}-10.2909 h^4 z^{25}  \nonumber\\
  &&-1700.11 h^3 z^{25}+139170. h^2 z^{25}+22.8001 h^4 z^{24}+1278.12 h^3 z^{24}-81413.6 h^2 z^{24}  \nonumber\\
  &&-17.4212 h^4 z^{23}-885.911 h^3 z^{23}+42312.9 h^2 z^{23}+8.18139 h^4 z^{22}+415.189 h^3 z^{22}  \nonumber\\
  &&-17696.9 h^2 z^{22}-2.92464 h^4 z^{21}-148.893 h^3 z^{21}+6113.61 h^2 z^{21}+0.859906 h^4 z^{20}  \nonumber\\
  &&+45.7754 h^3 z^{20}-1856.9 h^2 z^{20}-20.7656 h z^{20}-0.170264 h^4 z^{19}-9.83864 h^3 z^{19}  \nonumber\\
  &&+439.606 h^2 z^{19}+87.979 h z^{19}+0.0142505 h^4 z^{18}+1.36456 h^3 z^{18}-77.2301 h^2 z^{18}  \nonumber\\
  &&-119.517 h z^{18}-0.000939544 h^4 z^{17}-0.198901 h^3 z^{17}+6.62721 h^2 z^{17}+12.1008 h z^{17}  \nonumber\\
  &&+0.00398293 h^4 z^{16}+0.194368 h^3 z^{16}+5.9997 h^2 z^{16}+127.343 h z^{16}-0.000293095 h^4 z^{15}  \nonumber\\
  &&+0.0196677 h^3 z^{15}+0.227395 h^2 z^{15}-178.389 h z^{15}-0.00672851 h^4 z^{14}-0.384228 h^3 z^{14}  \nonumber\\
  &&-6.82685 h^2 z^{14}+200.178 h z^{14}+0.00706819 h^4 z^{13}+0.362624 h^3 z^{13}+6.14854 h^2 z^{13}  \nonumber\\
  &&-194.263 h z^{13}-0.00356212 h^4 z^{12}-0.143991 h^3 z^{12}-1.86424 h^2 z^{12}+117.789 h z^{12}  \nonumber\\
  &&+0.00122514 h^4 z^{11}+0.0438614 h^3 z^{11}+0.478409 h^2 z^{11}-41.6543 h z^{11}-0.000413594 h^4 z^{10}  \nonumber\\
  &&-0.0171749 h^3 z^{10}-0.230309 h^2 z^{10}+12.8821 h z^{10}+0.000100239 h^4 z^9+0.00227872 h^3 z^9  \nonumber\\
  &&-0.00174385 h^2 z^9-4.11834 h z^9-0.0000229442 h^4 z^8-0.00104187 h^3 z^8-0.0155285 h^2 z^8  \nonumber\\
  &&+0.626678 h z^8+4.89632\times 10^{-6} h^4 z^7+0.0000675029 h^3 z^7-0.00145586 h^2 z^7-0.227873 h z^7  \nonumber\\
  &&-7.4797\times 10^{-7} h^4 z^6-0.0000380442 h^3 z^6-0.000632556 h^2 z^6+0.0160514 h z^6  \nonumber\\
  &&+1.51157\times 10^{-7} h^4 z^5+1.29466\times 10^{-6} h^3 z^5-0.0000696617 h^2 z^5-0.00758056 h z^5  \nonumber\\
  &&-1.68373\times 10^{-8} h^4 z^4-9.41731\times 10^{-7} h^3 z^4-0.0000168844 h^2 z^4+0.00023328 h z^4  \nonumber\\
  &&+2.96525\times 10^{-9} h^4 z^3+1.74190\times 10^{-8} h^3 z^3-1.61648\times 10^{-6} h^2 z^3-0.00015552 h z^3  \nonumber\\
  &&-4.04039\times 10^{-10} h^4 z^2-2.25999\times 10^{-8} h^3 z^2-4.05296\times 10^{-7} h^2 z^2+3.1104\times 10^{-6} h z^2  \nonumber\\
  &&-z^2-0.0000186001 h^4 z-0.000305045 h^3 z-0.000671488 h^2 z+0.0277148 h z+z \label{solmuz}
\end{eqnarray}

 \vspace{1cm}
{\bf Acknowledgments}
 \vspace{1cm}

M. Z. is supported by Natural Science Basic Research Program of Shaanxi (Program No.2023-JC-QN-0053). D. C. Z acknowledges financial support from the Initial Research Foundation of Jiangxi Normal  University.

\end{document}